\documentclass[a4paper,fleqn]{cas-sc}

\usepackage[authoryear,longnamesfirst,numbers]{natbib}
\usepackage{amsmath,amsfonts}
\usepackage{url}
\usepackage{algpseudocode}
\usepackage{algorithm}
\usepackage{xcolor}
\usepackage{booktabs}
\usepackage{braket}
\usepackage{cleveref}
\crefname{equation}{}{}
\Crefname{equation}{Equation}{Equations}
\crefname{figure}{Fig.}{Figs.}
\crefname{table}{Table}{Tables}
\crefname{section}{Section}{Sections}
\crefname{algorithm}{Algorithm}{Algorithms}
\usepackage{float}
\usepackage{lineno}

\begin{document}

\shorttitle{AQPF and AQOPF Scalability Analysis}    
\shortauthors{Z. Kaseb et al.}  

\title [mode = title]{Power flow and optimal power flow using quantum and digital annealers: a computational scalability analysis}  

\author[1]{Zeynab Kaseb}[orcid=0000-0002-5142-290]

\cormark[1]
\ead{Z.Kaseb@tudelft.nl}

\author[2]{Matthias M\"oller}
\author[1]{Pedro P. Vergara}
\author[1]{Peter Palensky}

\affiliation[1]{organization={Electrical Sustainable Energy, Delft University of Technology, The Netherlands}}
\affiliation[2]{organization={Applied Mathematics, Delft University of Technology, The Netherlands}}
\cortext[1]{Corresponding author}

\begin{abstract}
This study further explores reformulating power flow (PF) analysis as a discrete combinatorial optimization problem, proposed in our earlier study using the Adiabatic Quantum Power Flow (AQPF) algorithm, which can be executed on Ising machines, including quantum and quantum-inspired hardware. This approach provides a new representation of the underlying equations, analogous to how neural networks approximate complex functions using simple operations. While the resulting combinatorial optimization problem is NP-hard, it is compatible with emerging quantum hardware designed to address such complexity. We introduce the Adiabatic Quantum Optimal Power Flow (AQOPF) algorithm, which transforms the classical optimal power flow (OPF) equations into quadratic unconstrained binary optimization (QUBO) models. Furthermore, the AQPF and AQOPF algorithms are evaluated on standard test cases ranging from 4- to 1354-bus systems using D-Wave's Advantage\texttrademark\ system (QA), its hybrid quantum-classical solver (HA), and Fujitsu's third-generation Digital Annealer (DAv3) and Quantum-Inspired Integrated Optimization (QIIO) platform. Both full and partitioned formulations are investigated, with particular attention to scalability and robustness in ill-conditioned scenarios. The results demonstrate that the algorithms can reproduce feasible PF and OPF solutions and exhibit promising computational scalability when supported by scalable hardware.
\end{abstract}



\maketitle


\section{Introduction}\label{}
Power flow (PF) and optimal power flow (OPF) are fundamental tasks in power system operation, planning, and control. PF analysis determines the steady-state voltages, power injections, and power flows in electrical grids under given operating conditions. OPF extends PF analysis by minimizing generation costs, reducing power losses, and/or improving voltage stability while satisfying operational constraints~\cite{Xiang2026,Jing2023}. With the increasing integration of renewable energy sources, demand-side management, and network congestion, efficient OPF solutions are critical to ensuring grid reliability~\cite{Li2025}. The Newton-Raphson (NR) method is widely used to solve PF and OPF problems due to its fast convergence rate and robustness for well-conditioned cases. It is an iterative numerical method that linearizes the PF equations using the Jacobian matrix and updates the solution until convergence is achieved. However, despite its efficiency in small- to medium-sized systems, NR faces significant computational challenges when applied to large-scale electrical grids with thousands of buses~\cite{Liu2022}. These challenges arise from repeated Jacobian matrix computations, matrix inversions, and high memory requirements, resulting in increased computational time and convergence issues. In addition, NR struggles with ill-conditioned power system scenarios, where numerical instability arises from poorly scaled Jacobian matrices. Such cases often arise in networks with high R/X ratios, weakly connected areas, or stressed operating conditions, for instance, during contingencies, voltage collapse scenarios, or low-inertia grid conditions~\cite{Nur2021}. These limitations make NR impractical for real-time operation, motivating the search for alternative methods~\cite{Liu2020_}.

Several approaches have been proposed in the literature to address the limitations of NR in large-scale, ill-conditioned power systems. For example, a fast NR-based PF algorithm is developed in~\cite{Ahmadi2022} that uses sparse matrix techniques and parallel processing to improve computational efficiency, particularly for large-scale power systems. Another example is~\cite{Costilla-Enriquez2021} where the authors propose a hybrid approach that integrates NR with stochastic gradient descent to enhance convergence properties, especially in ill-conditioned cases. In~\cite{Su2020}, a fully parallel NR implementation is also introduced that utilizes GPU-CPU vectorization and sparse techniques to accelerate PF analysis. Finally, an alternative PF controller is explored in ~\cite{Kumari2020} to improve PF capability, indirectly addressing some of the limitations of conventional NR-based methods. Note that while parallel NR implementations improve scalability, they still rely on matrix operations that become increasingly expensive as network sizes grow. Machine learning-based approaches, on the other hand, have been proposed to approximate PF and OPF solutions, but their accuracy depends on training data, limiting their generalization ability~\cite{Kaseb2026pinn,Liu2024,Wu2024}. These studies demonstrate that enhancing NR with advanced numerical techniques, hybrid approaches, and parallelization can significantly improve its applicability to large-scale power systems. Yet, fundamental challenges remain regarding computational complexity and convergence reliability. Thus, there is growing interest in the emerging field of quantum computing to address power system problems. Ising machines, in general, and quantum/digital annealing, in particular, offer the ability to efficiently explore large solution spaces, which makes them a promising candidate for tackling combinatorial optimization problems in power systems~\cite{Yin2023}. 

It is, however, important to emphasize that the aim of this study is not to replace the highly efficient classical algorithms developed for PF and OPF over several decades, e.g.,~\cite{Su2020,Ahmadi2022}. Classical numerical solvers remain the most practical and reliable tools for PF and OPF in current power system applications. Instead, this study further explores the alternative representation of the PF and OPF equations through a discrete combinatorial optimization approach proposed in our earlier study~\cite{kaseb2024power}. Similar to how neural networks approximate complex mathematical relationships using simple algebraic operations, the proposed reformulation discretizes the governing equations. This alternative representation provides a different computational perspective on PF and OPF, and can potentially address challenges encountered by numerical solvers in ill-conditioned scenarios. However, this representation introduces NP-hard optimization problems, which are generally challenging for classical algorithms but are compatible with emerging quantum hardware~\cite{Dent2026}. Although current quantum hardware remains limited in scale and precision, the development of algorithms that can operate on such devices is an important step toward understanding their potential future role in PF and OPF. The goal of this study is therefore to investigate whether PF and OPF can be reformulated in a way that may eventually enable quantum hardware to assist in addressing challenging scenarios that remain difficult for classical solvers, e.g., ill-conditioned networks.

That said, the present study extends the Adiabatic Quantum Power Flow (AQPF) algorithm, proposed in \cite{kaseb2024power}, to solve OPF problems, introducing the Adiabatic Quantum Optimal Power Flow (AQOPF) algorithm. The experiments are conducted on standard test cases ranging from 4- to 1354-bus systems, using D-Wave's Advantage\texttrademark\, system (QA), its hybrid quantum-classical solver (HA), as well as the third-generation Digital Annealer (DAv3) and Quantum-Inspired Integrated Optimization software (QIIO) developed by Fujitsu. The contributions of this paper are:
\begin{itemize}
    \item The AQPF algorithm, proposed in \cite{kaseb2024power}, is extended to solve OPF problems, introducing the AQOPF algorithm. AQPF is also evaluated using Fujitsu's latest QIIO software to assess its performance on a state-of-the-art quantum-inspired solver.
    \item A \emph{partitioned} formulation for AQPF and AQOPF is proposed, enabling partial problem-solving while maintaining accuracy comparable to the full formulations.  
    \item AQPF and AQOPF are shown to handle ill-conditioned cases where conventional solvers fail to converge effectively. The \emph{partitioned} formulation also exhibits similar robustness in handling such cases.  
    \item AQPF and AQOPF are shown to be computationally capable of addressing large-scale systems, provided the solver can handle a high number of binary variables.  
\end{itemize}

\section{Mathematical formulation} \label{sec:mathematical-formulation}
In a power system comprising $N$ buses, the relationship between the complex bus voltages and complex bus currents is described by Ohm's Law in matrix form, as $\mathbf{I} - \mathbf{Y}\mathbf{V} = 0$, where $\mathbf{I}$ and $\mathbf{V}$ represent the complex bus current and voltage vectors, respectively, and $\mathbf{Y}$ denotes the $N\times N$ complex bus admittance matrix, given by:
\begin{equation} \label{eq:addmittance}
    {Y}_{ik} = G_{ik} + jB_{ik}, \quad i,k\in\{1,\dots,N\},
\end{equation}
where $G_{ik}$ and $B_{ik}$ represent the conductance and susceptance between buses $i$ and $k$, respectively. We can further represent the net complex power flowing through the network as $\mathbf{S} = \mathbf{V}\mathbf{I}^* = \mathbf{V} \circ (\mathbf{Y}\mathbf{V})^*$, where $\mathbf{S} = \mathbf{P} + j\mathbf{Q}$ is a complex power injection vector and '$\circ$'~denotes element-wise vector multiplication. We can also express the complex power flowing through bus $i$ as the difference between the total generation power and the total demand power as:
\begin{equation} \label{eq:diff_power}
    {S}_i = {S}_i^G - {S}_i^D, \quad i\in\{1,\dots,N\},
\end{equation}
where $S_i$ denotes the net power flowing through bus $i$, $S_i^G$ and $S_i^D$ represent the total power generated and total power demand at bus $i$, respectively. Further expanding \cref{eq:diff_power} into its real and imaginary components gives:
\begin{equation} \label{eq:diff_power_split}
    P_i + jQ_i = (P_i^G - P_i^D) + j(Q_i^G - Q_i^D), \quad i\in\{1,\dots,N\}.
\end{equation}

Here, $P_i$ and $Q_i$ denote the net injections of active and reactive power at bus $i$; white $P_i^G$ and $Q_i^G$ correspond to the total generated power; and $P_i^D$ and $Q_i^D$ to the total power demand at that bus.

In this perspective, we can express the net active and reactive power injections at bus $i$ in terms of voltage and admittance parameters, as:
\begin{equation} \label{eq:pq_vi}
    P_i + jQ_i = {V}_i \sum_{k=1}^N {Y}_{ik}{V}_k, \quad i,k\in\{1,\dots,N\}.
\end{equation}

Further expanding \cref{eq:pq_vi} into real and imaginary components yields the AC PF equations in rectangular coordinates:
\begin{subequations} \label{eq:pq_sum}
    \begin{align}
        P_i & = \sum_{j=1}^N G_{ij}(\mu_i\mu_j + \omega_i\omega_j) + B_{ij}(\omega_i\mu_j - \mu_i\omega_j), \label{eq:p-sum} \\
        Q_i & = \sum_{j=1}^N G_{ij}(\omega_i\mu_j - \mu_i\omega_j) - B_{ij}(\mu_i\mu_j + \omega_i\omega_j) \label{eq:q-sum},
    \end{align}
\end{subequations}
where the complex voltage at bus $i\in\{1,\dots,N\}$ is expressed as ${V}_i=\mu_i+j\omega_i$, with $\mu_i$ and $\omega_i$ respectively representing its real and imaginary parts.

Each bus $i$ in a power system is defined by four primary variables: net active power injection ($P_i$), net reactive power injection ($Q_i$), voltage magnitude ($V_i$), and voltage angle ($\delta_i$). Based on the variables specified or controlled at each bus, buses in a power system are divided into three main types: \emph{PV} bus, \emph{PQ} bus, and \emph{slack} bus. 
\cref{tab:bus_types} summarizes the characteristics of these bus types. 
The \emph{slack} bus serves as a reference point for computing the PF and OPF.

\begin{table}[t!]
    \centering
    \caption{Types of buses in a power system.}
    \label{tab:bus_types}
    \begin{tabular}{cccc}
        \toprule
         & \emph{slack} bus & \emph{PV} bus & \emph{PQ} bus \\
        \midrule
        Known variables & $V_i$, $\delta_i$ & $P_i$, $V_i$ & $P_i$, $Q_i$ \\
        Unknown variables & $P_i$, $Q_i$ & $Q_i$, $\delta_i$ & $V_i$, $\delta_i$ \\
        Number of buses in a network & 1 & $N_G$ & $N - N_G - 1$ \\
        \bottomrule
    \end{tabular}
\end{table}

\subsection{Power Flow Formulation}
The objective of PF analysis is to compute the complex bus voltages and power injections that satisfy the power balance equations, as given in:
\begin{subequations} \label{eq:pf-cons}
    \begin{align}
        P_i & = P_i^G - P_i^D, \label{eq:active-power-balance}\\
        Q_i & = Q_i^G - Q_i^D.\label{eq:reactive-power-balance}
    \end{align}
\end{subequations}

These equations are iteratively solved, typically using numerical methods, such as the NR or Gauss-Seidel methods, to determine the complex bus voltages, i.e., $\mu_i$ and $\omega_i$ in rectangular form such that equations \cref{eq:pf-cons} hold for all buses $i\in\{1,\dots,N\}$. 

\subsection{Optimal Power Flow Formulation}
The OPF problem aims to minimize total generation costs while ensuring that all system constraints are satisfied. The goal is to compute the optimal generator outputs $\{P_i^G, Q_i^G\}$ and bus voltages $\{V_i, \delta_i\}$ that minimize generation costs while satisfying the power balance, generation, and operational limits. The problem is formulated as:
\begin{equation} \label{eq:opf-obj}
    \min \sum_{k \in \mathbb{G}}^{N_G} f_k\left(P_k^G\right),
\end{equation}
subject to
\begin{subequations} \label{eq:opf-cons}
    \begin{align}
        P_i & = P_i^G - P_i^D, \label{eq:opf-active-power-balance}\\
        Q_i & = Q_i^G - Q_i^D, \label{eq:opf-reactive-power-balance}\\
        \underline{P_i^G} & \leq P_i^G \leq \overline{P_i^G}, \label{eq:active-power-limits}\\
        \underline{Q_i^G} & \leq Q_i^G \leq \overline{Q_i^G}, \label{eq:reactive-power-limits}\\
        \underline{V_i} & \leq V_i \leq \overline{V_i}, \label{eq:voltage-limits}\\
        \underline{\delta_i} & \leq \delta_i \leq \overline{\delta_i}, \label{eq:angle-limits}
    \end{align}
\end{subequations}
where $f_k(.)$ is the generator's fuel cost function. $k \in \mathbb{G}$ is a subset of buses connected to generators \{1, 2, ..., $N_G$\}, where $N_G$ is the total number of generators. $\underline{P_i^G}$ and $\overline{P_i^G}$ are the minimum and maximum generated active power outputs at bus $i$, respectively; $\underline{Q_i^G}$ and $\overline{Q_i^G}$ are the minimum and maximum generated reactive power outputs at bus $i$, respectively; $\underline{V_i}$ and $\overline{V_i}$ are the minimum and maximum voltage magnitudes at bus $i$, respectively; $\underline{\delta_i}$ and $\overline{\delta_i}$ are the minimum and maximum voltage phase angles at bus $i$, respectively.

\section{Combinatorial formulation} \label{sec:combinatorial-formulation}
The classical PF and OPF are reformulated and restructured to obtain a combinatorial optimization problem that can be solved using Ising machines, e.g., quantum/digital annealers.

\subsection{Combinatorial Power Flow Formulation}
The discrepancy between the specified active and reactive power demands, $P_i^D$ and $Q_i^D$, active and reactive power injection, $P_i^G$ and $Q_i^G$, and their net counterparts, $P_i$ and $Q_i$, at bus $i\in\{1,\dots,N\}$ can be written as:
\begin{subequations} \label{eq:pq-mismatch}
    \begin{align}
        P_i - P_i^G + P_i^D &\stackrel{!}{=} 0, \label{eq:p-mismatch}\\
        Q_i - Q_i^G + Q_i^D &\stackrel{!}{=} 0, \label{eq:q-mismatch}
    \end{align}
\end{subequations}
where the goal is to minimize this mismatch to satisfy the power balance equations \eqref{eq:pf-cons}. To further convert the problem into a form that annealers can handle, we expand \cref{eq:pq_sum} into:
\begin{subequations} \label{eq:pq-sum_expanded}
    \begin{align}
        P_i & = \sum_{j=1}^{N} \mu_i G_{ij} \mu_j + \omega_i G_{ij} \omega_j + \omega_i B_{ij} \mu_j - \mu_i B_{ij} \omega_j,  \label{eq:p-sum_expanded}\\ 
        Q_i & = \sum_{j=1}^{N} \omega_i G_{ij} \mu_j - \mu_i G_{ij} \omega_j - \mu_i B_{ij} \mu_j - \omega_i B_{ij} \omega_j. \label{eq:q-sum_expanded}
    \end{align}
\end{subequations}

Here, the variables $\mu_i$, $\mu_k$, $\omega_i$, and $\omega_k$ are continuous real numbers that must be discretized to reformulate the problem as a \emph{binary} combinatorial optimization task. This leads naturally to a Quadratic Unconstrained Binary Optimization (QUBO) model, which is defined by its symmetric, real-valued matrix $Q\in\mathbb{R}^{n\times n}$ and the binary minimization problem:
\begin{equation}
    \min_{\mathbf{x} \in \{0,1\}^n} \sum_{i=1}^{n} \sum_{j=1}^{n} Q_{ij} x_i x_j. \label{eq:qubo}
\end{equation}

A basic discretization strategy for $\mu_i$ and $\omega_i$ involves encoding each as a base value offset by scaled binary variables:
\begin{subequations} \label{eq:muomega-increment}
\begin{align}
    \mu_i & = \mu_i^0 + \Delta \mu_i ( x_{i,0}^\mu - x_{i,1}^\mu ), \label{eq:mu-increment}\\
    \omega_i & = \omega_i^0 + \Delta \omega_i (x_{i,0}^\omega - x_{i,1}^\omega ), \label{eq:omega-increment}
\end{align}
\end{subequations}
where the variables $x_{i,\{0,1\}}^{\{\mu, \omega\}}\in\{0,1\}$ are \emph{binary} indicators that control deviations from the reference values $\mu_i^0$ and $\omega_i^0$. An increase occurs when $x_{i,0} = 1$ and $x_{i,1} = 0$, a decreased when $x_{i,0} = 0$ and $x_{i,1} = 1$, and no change if both bits are equal. While the case $\{00\}$ and $\{11\}$ introduce redundancy in representing no change, allowing all four bit combinations $\{00, 01, 10, 11\}$ maintains a complete and unconstrained binary encoding space. 

By inserting \cref{eq:muomega-increment} into \cref{eq:p-sum_expanded}, we derive an expression for the net active power injection $P_i$. Note that this expression follows the derivation in our previous work~\cite{kaseb2024power}, and is reproduced here for completeness:
\begin{equation} 
\label{eq:p-qubo}
\begin{split} 
    P_i=
    \sum_{j=0}^{N}
    \Big[ \mu_i^0 G_{ij} \mu_j^0
    + 
    \omega_i^0 G_{ij} \omega_j^0
    + 
    \omega_i^0 B_{ij} \mu_j^0
    - 
    \mu_i^0 B_{ij} \omega_j^0 \Big] \\
    +
    \sum_{j=0}^{N}\sum_{k=0}^{1} 
    \Big[ (-1)^k \mu_i^0 G_{ij} x_{j,k}^\mu\Delta \mu_j
    +
    (-1)^kx_{i,k}^\mu\Delta \mu_i G_{ij} \mu_j^0\\
    +
    (-1)^k\omega_i^0 G_{ij} x_{j,k}^\omega\Delta \omega_j
    +
    (-1)^kx_{i,k}^\omega\Delta \omega_i G_{ij} \omega_j^0\\[2ex]
    +
    (-1)^k\omega_i^0 B_{ij} x_{j,k}^\mu\Delta \mu_j
    + 
    (-1)^kx_{i,k}^\omega\Delta \omega_i B_{ij} \mu_j^0\\[2ex]
    -
    (-1)^k\mu_i^0 B_{ij} x_{j,k}^\omega\Delta \omega_j
    - 
    (-1)^kx_{i,k}^\mu\Delta \mu_i B_{ij} \omega_j^0 \Big] \\[2ex]
    +
    \sum_{j=0}^{N}\sum_{k=0}^{1}\sum_{l=0}^{1} 
    \Big[ (-1)^{k+l}x_{i,k}^\mu\Delta \mu_i G_{ij} x_{j,l}^\mu\Delta \mu_j\\[2ex]
    + 
    (-1)^{k+l}x_{i,k}^\omega\Delta \omega_i G_{ij} x_{j,l}^\omega\Delta \omega_j\\[2ex]
    + 
    (-1)^{k+l}x_{i,k}^\omega\Delta \omega_i B_{ij} x_{j,l}^\mu\Delta \mu_j\\[2ex]
    - 
    (-1)^{k+l}x_{i,k}^\mu\Delta \mu_i B_{ij} x_{j,l}^\omega\Delta \omega_j \Big].
\end{split}
\end{equation}


An analogous expression for the net reactive power $Q_i$ is obtained by inserting \eqref{eq:muomega-increment} into \eqref{eq:q-sum_expanded}. This expression is also based on our earlier work~\cite{kaseb2024power}:
\begin{equation} 
\label{eq:q-qubo}
\begin{split}
    Q_i=
    \sum_{j=0}^{N}
    \Big[\omega_i^0 G_{ij} \mu_j^0
    - 
    \mu_i^0 G_{ij} \omega_j^0
    - 
    \mu_i^0 B_{ij} \mu_j^0
    - 
    \omega_i^0 B_{ij} \omega_j^0\Big] \\
    +
    \sum_{j=0}^{N}\sum_{k=0}^{1}
    \Big[(-1)^k\omega_i^0 G_{ij} x_{j,k}^\mu\Delta \mu_j
    + 
    (-1)^kx_{i,k}^\omega\Delta \omega_i G_{ij} \mu_j^0\\
    -
    (-1)^k\mu_i^0 G_{ij} x_{j,k}^\omega\Delta \omega_j
    - 
    (-1)^kx_{i,k}^\mu\Delta \mu_i G_{ij} \omega_j^0\\[2ex]
    - 
    (-1)^k\mu_i^0 B_{ij} x_{j,k}^\mu\Delta \mu_j
    - 
    (-1)^kx_{i,k}^\mu\Delta \mu_i B_{ij} \mu_j^0\\[2ex]
    -
    (-1)^k\omega_i^0 B_{ij} x_{j,k}^\omega\Delta \omega_j
    - 
    (-1)^kx_{i,k}^\omega\Delta \omega_i B_{ij} \omega_j^0\Big]\\[2ex]
    +
    \sum_{j=0}^{N}\sum_{k=0}^{1}\sum_{l=0}^{1}
    \Big[(-1)^{k+l}x_{i,k}^\omega\Delta \omega_i G_{ij} x_{j,l}^\mu\Delta \mu_j\\[2ex]
    - 
    (-1)^{k+l}x_{i,k}^\mu\Delta \mu_i G_{ij} x_{j,l}^\omega\Delta \omega_j\\[2ex]
    - 
    (-1)^{k+l}x_{i,k}^\mu\Delta \mu_i B_{ij} x_{j,l}^\mu\Delta \mu_j\\[2ex]
    - 
    (-1)^{k+l}x_{i,k}^\omega\Delta \omega_i B_{ij} x_{j,l}^\omega\Delta \omega_j\Big].
\end{split}
\end{equation}

The ansatz presented in \cref{eq:p-qubo,eq:q-qubo} yields a QUBO structure, as expressed in \eqref{eq:qubo}, enabling the minimization of the squared mismatch terms in \cref{eq:pq-mismatch} using annealing approaches:
\begin{align} \label{eq:H_obj}
    H_{\text{obj}}(\mathbf{x}) = \sum\limits_{i=1}^{N} (P_i - P_i^G + P_i^D)^2 + (Q_i - Q_i^G + Q_i^D)^2,
\end{align}
where $\mathbf{x}$ represents the vector of binary variables.

\subsection{Combinatorial Optimal Power Flow Formulation}
In contrast to the previously described PF analysis, power balance equations \eqref{eq:pf-cons} become equality constraints in the \mbox{\emph{optimal}} PF problem, i.e., \cref{eq:opf-active-power-balance,eq:opf-reactive-power-balance}, and, consequently, \eqref{eq:H_obj} becomes a penalty term in the QUBO formulation of the OPF problem, $\lambda_{\text{PF}}H_{\text{obj}}(\cdot)$, where $\lambda_{\text{PF}}$ is the corresponding penalty parameter. Another penalty term is needed to implement the inequality constraint terms, $H_{\text{const}}(\cdot)$, as well as the objective function of the OPF problem as an equality constraint term, $H_{\text{cost}}(\cdot)$. Let us start by describing the Lagrange multiplier approach, which imposes equality constraints in a weak sense within a QUBO formulation. To this end, consider the following minimization problem with a single equality constraint:
\begin{subequations} \label{eq:ineq-const}
    \begin{align} 
        \min_x \quad f(x)\\
        s.t. \quad g(x) = c .
    \end{align} 
\end{subequations}
An equivalent formulation of the problem reads as follows:
\begin{equation} \label{eq:equality_penalty}
        \min_x \quad f(x) + \lambda \left(g(x)-c\right)^2, \quad\lambda>0,
\end{equation}
where $\lambda$ is a penalty parameter that ensures that any violation of the equality constraint incurs a cost in the QUBO formulation. A set of $m$ equality constraints on $n$ primary variables can be written in matrix form as follows:
\begin{equation} \label{eq:linear_eq_constraints}
\mathbf{C} \mathbf{x} = \mathbf{c}, \quad \mathbf{C} \in \mathbb{R}^{m \times n}, \quad \mathbf{c} \in \mathbb{R}^{m},
\end{equation}
and imposed via the Lagrange multiplier approach as:
\begin{equation} \label{eq:qubo_eq_constraints}
P(\mathbf{x}) = \sum_{i=1}^{m} \lambda_i \left(\sum_{j=1}^nC_{ij} x_j - c_i\right)^2, \quad \lambda_i > 0.
\end{equation}

Next, we describe the procedure for converting inequality constraints into equality constraints by adopting slack variables. To this end, consider the general minimization problem with a single inequality constraint:
\begin{subequations} \label{eq:eq-const}
    \begin{align} 
        \min_x \quad f(x)\\
        s.t. \quad h(x) \leq 0,
    \end{align} 
\end{subequations}
and introduce the non-negative slack variable $s\in\mathbb{R}_{\ge 0}$ to obtain an optimization problem with equality constraint:
\begin{subequations} \label{eq:eq-const-slack}
    \begin{align}
        \min_x \quad f(x)\\
        s.t. \quad h(x) + s = 0 \quad s \geq 0.
    \end{align}
\end{subequations}
To enforce \(m\) inequality constraints on \(n\) binary variables, 
\begin{equation} \label{eq:ineq_constraint}
\mathbf{D} \mathbf{x} \leq \mathbf{d}, \quad \mathbf{D} \in \mathbb{R}^{m \times n}, \quad \mathbf{d} \in \mathbb{R}^{m},
\end{equation}
we introduce a vector of slack variables $\mathbf{s}\in\mathbb{R}^{n}_{\ge 0}$ such that
\begin{equation} \label{eq:slack_variable}
\mathbf{D} \mathbf{x} + \mathbf{s} = \mathbf{d}, \quad \mathbf{s} \geq \mathbf{0}.
\end{equation}

Since QUBO formulations require binary variables, we represent each slack variable using binary encoding
\begin{equation} \label{eq:binary_slack}
s_i = \sum_{j=0}^{k} 2^j a_{i,j}, \quad a_{i,j} \in \{0,1\},
\end{equation}
where $a_{i,j}$ are binary variables representing the discretized values of the slack variable $s_i$. By substituting expression \eqref{eq:binary_slack} into \cref{eq:slack_variable}, the original inequality constraints can be transformed into an equivalent QUBO formulation:
\begin{equation} \label{eq:qubo_ineq_constraints}
P(\mathbf{x}, \mathbf{a}) = \sum_{i=1}^{m} \lambda_i \left(\sum_{j=1}^{n}D_{ij} x_j + s_i - d_i\right)^2, \quad \lambda_i > 0.
\end{equation}

Utilizing the slack variable approach, the inequality constraints \eqref{eq:active-power-limits}--\eqref{eq:angle-limits} can be converted into equality constraints:
\begin{subequations} \label{eq:ineq-constraints-slack}
\begin{align}
P_i^G = \underline{P_i^G} + \Delta P \cdot s_{P_i^+}, \quad s_{P_i^+} \geq 0,\\
P_i^G = \overline{P_i^G} - \Delta P \cdot s_{P_i^-}, \quad s_{P_i^-} \geq 0, \\
Q_i^G = \underline{Q_i^G} + \Delta Q \cdot s_{Q_i^+}, \quad s_{Q_i^+} \geq 0,\\
Q_i^G = \overline{Q_i^G} - \Delta Q \cdot s_{Q_i^-}, \quad s_{Q_i^-} \geq 0, \\
V_i = \underline{V_i} + \Delta V \cdot s_{V_i^+}, \quad s_{V_i^+} \geq 0,\\
V_i = \overline{V_i} - \Delta V \cdot s_{V_i^-}, \quad s_{V_i^-} \geq 0, \\
\delta_i = \underline{\delta_i} + \Delta \delta \cdot s_{\delta_i^+}, \quad s_{\delta_i^+} \geq 0,\\
\delta_i = \overline{\delta_i} - \Delta \delta \cdot s_{\delta_i^-}, \quad s_{\delta_i^-} \geq 0,
\end{align}
\end{subequations}
which can be imposed as penalty terms to ensure that the active and reactive power generation limits, voltage limits, and angle limits are respected:
\begin{subequations} \label{eq:qubo_ineq_constraints}
\begin{align}
H_{\text{const}}(\mathbf{x}) = \sum_{i} \lambda_{P_i^+} \left(P_i^G - \underline{P_i^G} - \Delta P \cdot s_{P_i^+}\right)^2 \\
+ \lambda_{P_i^-} \left(\overline{P_i^G} - P_i^G - \Delta P \cdot s_{P_i^-}\right)^2 \\
+ \lambda_{P_i} \left(\underline{P_i^G} - \overline{P_i^G} + \Delta P \cdot s_{P_i^+} - \Delta P \cdot s_{P_i^-}\right)^2 \\
\lambda_{Q_i^+} \left(Q_i^G - \underline{Q_i^G} - \Delta Q \cdot s_{Q_i^+}\right)^2 \\
+ \lambda_{Q_i^-} \left(\overline{Q_i^G} - Q_i^G - \Delta Q \cdot s_{Q_i^-}\right)^2 \\
+ \lambda_{Q_i} \left(\underline{Q_i^G} - \overline{Q_i^G} + \Delta Q \cdot s_{Q_i^+} - \Delta Q \cdot s_{QP_i^-}\right)^2 \\
\lambda_{V_i^+} \left(V_i - \underline{V_i} - \Delta V \cdot s_{V_i^+}\right)^2 \\
+ \lambda_{V_i^-} \left(\overline{V_i} - V_i - \Delta V \cdot s_{V_i^-}\right)^2 \\
+ \lambda_{V_i} \left(\underline{V_i} - \overline{V_i} + \Delta V \cdot s_{V_i^+} - \Delta V \cdot s_{V_i^-}\right)^2 \\
\lambda_{\delta_i^+} \left(\delta_i - \underline{\delta_i} - \Delta \delta \cdot s_{\delta_i^+}\right)^2 \\
+ \lambda_{\delta_i^-} \left(\overline{\delta_i} - \delta_i - \Delta \delta \cdot s_{\delta_i^-}\right)^2 \\
+ \lambda_{\delta_i} \left(\underline{\delta_i} - \overline{\delta_i} + \Delta \delta \cdot s_{\delta_i^+} - \Delta \delta \cdot s_{\delta_i^-}\right)^2.
\end{align}
\end{subequations}

The actual cost function of the Hamiltonian is defined as:
\begin{equation} \label{eq:eq_const}
    H_{\text{cost}}(\mathbf{x}) = \lambda_{P_k^G} \sum_{k \in \mathbb{G}}^{N_G} f_k\left(P_k^G\right)^2.
\end{equation}

Combining all three components yields the following problem Hamiltonian $H(\cdot)$:
\begin{equation}
    \label{eq:final-hamiltonian}
    H(\mathbf{x}) = \lambda_{\text{PF}} H_{\text{obj}}(\mathbf{x}) + H_{\text{const}}(\mathbf{x}) + H_{\text{cost}}(\mathbf{x}).
\end{equation}

\section{Quantum and Digital Annealing} \label{sec:annealing}
Quantum annealing is an analog computing paradigm designed to solve unconstrained optimization problems formulated as an Ising models~\cite{Goto2019}. The Ising Hamiltonian reads:
\begin{equation} \label{eq:ising}
    \min_{\vec{s}\in\{\pm 1\}^n} \sum_{\langle i,j\rangle} J_{ij}s_is_j + \sum_{i=1}^n h_i s_i,
\end{equation}
where $J_{ij}$ represents interaction coefficients between spins, and $h_i$ represents an external field, which acts as a linear bias that influences the tendency of \(s_i\) toward \(+1\) and \(-1\); $\langle i,j\rangle$ represents all connections between $i,j\in\{1,\dots,n\}$ such that $i<j$. The system follows the dynamics of the time-dependent Schr{\"o}dinger equation, starting from a highly delocalized quantum state and evolves according to a time-varying Hamiltonian. This evolution is designed such that, under ideal conditions, the system settles into the ground state of the final Hamiltonian, which corresponds to the optimal solution \cite{Lucas2014}. The QUBO formulations derived in \cref{sec:combinatorial-formulation} can be converted into Ising models by setting $x_i=(s_i-1)/2$ and regrouping the coefficients into $J_{ij}$ and $h_i$.

Quantum and Digital Annealer (DA) are specialized hardware architectures designed to minimize the energy of quadratic functions over binary variables using this optimization approach. These architectures are referred to as Ising machines~\cite{Bunyk2014,Goto2021}. D-Wave's Advantage\texttrademark\ system\footnote{\url{https://www.dwavesys.com}} is one of the most well-known implementations of quantum annealing, featuring over 5,000 qubits and 35,000 couplers. In this system, a positive weight $J_{ij}>0$ encourages opposite spin orientations (ferromagnetic interaction), while a negative weight $J_{ij}<0$ favors aligned spins (antiferromagnetic interaction). Similarly, $h_i>0$ implies that the spin at site $i$ favors lining up in the positive direction, whereas $h_i<0$ implies the opposite. Practical challenges arise due to the limited qubit connectivity, necessitating minor embedding to map the problem graph onto the hardware topology, an NP-hard problem. In addition, the range of $J$ and $h$ values is restricted, thus leading to a potential loss of precision in problem encoding.

To overcome these challenges, Fujitsu introduced the DA, an application-specific complementary metal-oxide semiconductor (CMOS) hardware\footnote{\url{https://www.fujitsu.com/global/services/business-services/digital-annealer}} that emulates simulated annealing. It supports massively parallel execution of Markov Chain Monte Carlo (MCMC) processes and offers significantly higher qubit connectivity than quantum annealing. Fujitsu's DA Unit can handle up to 100,000 fully connected binary variables with 64-bit precision. Unlike quantum annealing, which, depending on the qubit technology used, requires cryogenic temperatures, digital annealing operates at room temperature, making it suitable for broader deployment, including edge computing applications. Each run of DA begins with a randomly generated binary configuration and iteratively explores the solution space to identify a state that yields minimal energy. Another key advantage of digital annealing over quantum annealing is its native support for QUBO problems~\cite{Inoue2024}. Building on simulated annealing, the second-generation DA uses parallel tempering. This method simulates multiple solution paths at different temperature levels and enables configuration swaps between them to improve exploration and prevent stagnation in local minima.

A significant limitation of both quantum and digital annealing, however, is their restriction to binary quadratic cost functions. Many real-world problems, such as OPF, require handling higher-order polynomial cost functions and inequality constraints. Recent advancements in DA introduce support for higher-order terms and constrained optimization. DA achieves this by employing various operations to reduce the computational cost of multiplications and by applying replica-exchange Monte Carlo methods to enhance search efficiency. These improvements significantly expand the range of applications beyond traditional QUBO problems, including large-scale power systems. The software implementation of \cite{Yin2023} is available as QIIO\footnote{\url{https://en-portal.research.global.fujitsu.com/kozuchi}}, the name for the latest DA prototype.

\subsection{Higher-Order Terms Handling}
Expanding the full expressions of \cref{eq:H_obj} for PF and \cref{eq:final-hamiltonian} for OPF results in a fourth-degree polynomial involving $\mathbf{x}$. QIIO supports native optimization over higher-order polynomials, but DAv3, QA, and HA are limited to quadratic terms. Consequently, higher-order interactions must be reformulated into equivalent quadratic representations to be compatible with these solvers. One standard method for reducing cubic terms is to introduce auxiliary variables, such as $z_{ij}=x_i x_j$, allowing triplet products to be reformulated as:
\begin{align} \label{eq:three-terms}
    x_i x_j x_k = \min_{z_{ij}} \left\{z_{ij}x_k + \lambda P(x_i, x_j; z_{ij})\right\},\quad \lambda > 0,
\end{align} 
where the penalty function $P$ enforces the consistency of $z_{ij}$ with $x_i$ and $x_j$ and is given by:
\begin{align} \label{eq:penalty-P}
    P(x_i, x_j; z_{ij}) = x_i x_j - 2(x_i+x_j)z_{ij} + 3z_{ij}.
\end{align}

Similarly, quartic terms involving four binary variables can be reduced using two auxiliary variables and the penalty formulation as follows:
\begin{align} \label{eq:four-terms}
    x_i x_j x_k x_l = \min_{z_{ij}, z_{kl}} \left\{\lambda P(x_i, x_j; z_{ij}) + \lambda P(x_k, x_l; z_{kl})\right\}.
\end{align}

For an in-depth review of quadratization approaches, see \cite{Dattani2019}. To construct the QUBO formulations corresponding to \cref{eq:H_obj} for PF and \cref{eq:final-hamiltonian} for OPF, while also reducing higher-order terms, we employ the PyQUBO package for QA and HA solvers, and the Python package DADK for DAv3. After solving the minimization problem, the resulting bitstring is used to update $\mu_i$ and $\omega_i$ according to \cref{eq:muomega-increment}. Note that we neglect the auxiliary variables $z_{ij}$ for the bus complex voltage updates. 

Our experiments reveal discrepancies between the PyQUBO and DADK implementations, particularly in the number of auxiliary variables generated when encoding problems of identical size. In general, DADK tends to produce more compact formulations with fewer auxiliary variables. We verify that the order reduction preserves the optimality of solutions and maintains consistency in variable assignments corresponding to the minimum-energy configurations.

\section{Adiabatic Quantum Algorithm} \label{sec:adiabatic-algorithms}
Iterative schemes are introduced to ensure that AQPF and AQOPF use the unique computational properties of quantum/digital annealers to perform combinatorial PF and OPF, respectively, as presented in \cref{alg:aqpf}. Initially, the active power generation vector $\mathbf{P}^{G}$, the demand vectors $\mathbf{P}^{D}$ and $\mathbf{Q}^{D}$, along with the power system admittance matrix $\mathbf{Y}$, are defined (lines 1-4). Next, the voltage real and imaginary parts, $\mathbf{\mu}$ and $\mathbf{\omega}$, as well as their corresponding adjustments, $\Delta{\mu}$ and $\Delta{\omega}$, are initialized by values provided by the user (refer to lines 5–8). These parameters are then used to compute the starting values of the active and reactive power vectors, $\mathbf{P}$ and $\mathbf{Q}$ (see lines 9–10). Note that $P_1$ and $Q_1$, associated with the \emph{slack} bus, are excluded to align with the PF problem. 

On line 11, the Hamiltonian $H_\text{obj}(\cdot)$ is evaluated at the binary vector $\mathbf{x}=[x_{0,0}^\mu,\dots,x_{n,1}^\omega] \in \{0\}^{4n}$ (line 11) using the initial voltage estimations to propose a candidate solution. In line 12, the residual limit, defined as $\epsilon$ ${\text{MW}^2 + \text{MVAR}^2}\over 2$, is initialized, followed by resetting the iteration counter `$\text{it}$' to zero before entering the iterative loop at line 13. After solving the minimization problem \cref{eq:H_obj} (line 15), the binary solution vector $\mathbf{x}=[x_{0,0}^\mu,\dots,x_{n,1}^\omega] \in \{0,1\}^{4n}$ derived from the QUBO formulation is employed to revise $\mu_i$ and $\omega_i$ as specified in line 16 by \cref{eq:muomega-increment}. It is important to note that during the voltage update process, any auxiliary variables introduced through the quadratic reformulation of higher-order terms are intentionally omitted. When the computed $H_\text{obj}(\mathbf{x})$ falls below the user-defined threshold $\epsilon$, the resulting complex voltages $\mathbf{\mu}+j\mathbf{\omega}$ are considered as the final solution. If not, the base values, $\mu_i^0$ and $\omega_i^0$, are reassigned from the current $\mu_i$ and $\omega_i$ (line 20), and the optimization routine for \cref{eq:H_obj} is repeated until convergence is achieved. 


For AQOPF, \cref{alg:aqpf} should be modified to align with the OPF problem principle. First, line 1 is not needed, as the generator set points are determined by solving the OPF problem. Moreover, the active and reactive power vectors, $\mathbf{P}$ and $\mathbf{Q}$, are computed only for load buses in lines 9, 10, 17, and 18. Finally, the problem Hamiltonian in lines 11 and 18 should be replaced with \cref{eq:final-hamiltonian}. 

\begin{algorithm}[t!]
\caption{Adiabatic quantum algorithm for PF analysis.}\label{alg:aqpf}
\begin{algorithmic}[1]
    \State Initialize $\mathbf{P}^{G} = [P^{G}_1, P^{G}_2, \dots, P^{G}_{N_G}]$
    \State Initialize $\mathbf{P}^{D} = [P^{D}_1, P^{D}_2, \dots, P^{D}_{N - N_G - 1}]$
    \State Initialize $\mathbf{Q}^{D} = [Q^{D}_1, Q^{D}_2, \dots, Q^{D}_{N - N_G - 1}]$
    \State Initialize $\mathbf{Y}= \{(G_{ij}+\textrm{j}B_{ij}): i,j=1,2,\dots,N\}$
    \State $\Delta{\mu} \gets 1 \times 10^{-2}$
    \State $\Delta{\omega} \gets 1 \times 10^{-3}$ 
    \State $\mathbf{\mu}^0 = [\mu_{1}^0, \mu_{2}^0, \dots, \mu_{N}^0] \gets 1$
    \State $\mathbf{\omega}^0 = [\omega_{1}^0, \omega_{2}^0, \dots, \omega_{N}^0] \gets 0$
    \State Calculate $\mathbf{P} = [P_2, P_3, \dots, P_{N}]$ using \cref{eq:p-qubo}
    \State Calculate $\mathbf{Q} = [Q_2, Q_3, \dots, Q_{N}]$ using \cref{eq:q-qubo}
    \State Calculate Hamiltonian $H_{\text{obj}}(\mathbf{x})$ using \cref{eq:H_obj}
    \State $\epsilon \gets 1 \times 10^{-2}$
    \State $\text{it} \gets 0$
    \While{$H_{obj}(\mathbf{x}) > \epsilon$ and $\text{it}<\text{it}_{\text{max}}$}
        \State Update $\mathbf{x}$
        \State Update $\mathbf{\mu}$ and $\mathbf{\omega}$ using \cref{eq:muomega-increment}
        \State Recalculate $\mathbf{P}$ using \cref{eq:p-qubo}
        \State Recalculate $\mathbf{Q}$ using \cref{eq:q-qubo}
        \State Recalculate $H_{\text{obj}}(\mathbf{x})$ using \cref{eq:H_obj}
        \State Update $\mathbf{\mu}^0$ and $\mathbf{\omega}^0$ 
        \State Update $\Delta{\mu}$, $\Delta{\omega}$
        \State $\text{it} \gets \text{it} + 1$
    \EndWhile
\end{algorithmic}
\end{algorithm}

\subsection{Delta Update}
The iterative method may exhibit slow convergence when $\Delta \mu_i$ and $\Delta \omega_i$ are set too low, or if $\mu_i^0$ and $\omega_i^0$ are significantly different from the values that fulfill \cref{eq:H_obj,eq:final-hamiltonian}. A possible remedy is to employ an adaptive strategy for $\Delta \mu_i$ and $\Delta \omega_i$, dynamically adjusting their values at each iteration and for each bus $i$ to enhance convergence. In this perspective, the values for $\Delta \mu_i$ and $\Delta \omega_i$ are a function of the iteration counter `$\text{it}$', as given in:
\begin{subequations} \label{eq:delta-muomega-update}
\begin{align}
    \Delta \mu_i 
    &=
    \underline{\Delta \mu_i} + (\overline{\Delta \mu_i} - \underline{\Delta \mu_i}) \times e^{r \cdot \text{it}}, \label{eq:delta-mu-update}\\[2ex]
    \Delta \omega_i
    &=
    \underline{\Delta \omega_i} + (\overline{\Delta \omega_i} - \underline{\Delta \omega_i}) \times e^{r \cdot \text{it}}, \label{eq:delta-omega-update}
\end{align}
\end{subequations}
where $\underline{\Delta \mu_i}=5 \times 10^{-4}$ and $\overline{\Delta \mu_i}=4 \times 10^{-2}$ are lower and upper bounds of $\Delta \mu_i$, respectively; Similarly, $\underline{\Delta \omega_i}=1 \times 10^{-4}$ and $\overline{\Delta \omega_i}=2 \times 10^{-2}$ bound $\Delta \omega_i$; $e$ is the exponential function; $r=-0.05$ is the rate of decay. In each iteration, $\Delta\mathbf{\mu}$ and $\Delta\mathbf{\omega}$ are updated adaptively and individually; we save two sets of binary variables $\mathbf{x}^{(\text{it}-1)}$ and $\mathbf{x}^{(\text{it}-2)} \in \{0,1\}^{4n}$, which represent the bitstrings for the previous and second previous iterations, respectively. We then use these bitstrings to update $\Delta \mu_i$ and $\Delta \omega_i$ for each bus $i$. For example, $x_{i,0}^{\mu(\text{it}-2)} = 1 \land x_{i,1}^{\mu(\text{it}-2)} = 0$, $x_{i,0}^{\mu(\text{it}-1)} = 1 \,\land\, x_{i,1}^{\mu(\text{it}-1)} = 0$, and $x_{i,0}^\mu = 0 \land x_{i,1}^\mu = 1$ indicates that at iteration $\text{it}-2$, the value of $\mu_i$ increased, followed by an increase at iteration $\text{it}-1$ and then a decrease at the current iteration `$\text{it}$'. This pattern suggests that to prevent $\mu_i$ from fluctuations and promote convergence, we should reduce $\Delta \mu_i$. Details are given in \cref{alg:qubo-update-deltas}. We employ a similar approach to update $\Delta \mathbf{\omega}$.

\begin{algorithm}[t!]
\caption{$\Delta \boldsymbol{\mu}$ update algorithm for the QUBO formulation.} \label{alg:qubo-update-deltas} 
\begin{algorithmic}[1]
\For{$\text{it}=0 \to \text{it}_{\max}$}
    \For{$i = 0 \to \text{len}(\Delta\boldsymbol{\mu})$}
        \If{$x_{i,0}^\mu + x_{i,0}^{\mu(\text{it}-1)} + x_{i,0}^{\mu(\text{it}-2)} + x_{i,1}^\mu + x_{i,1}^{\mu(\text{it}-1)} + x_{i,1}^{\mu(\text{it}-2)} = 0 \lor x_{i,0}^\mu + x_{i,0}^{\mu(\text{it}-1)} + x_{i,0}^{\mu(\text{it}-2)} + x_{i,1}^\mu + x_{i,1}^{\mu(\text{it}-1)} + x_{i,1}^{\mu(\text{it}-2)} = 6$}
            \State $\Delta \mu_i \gets \Delta \mu_i/2$
        \ElsIf{$x_{i,0}^\mu = 0 \land x_{i,0}^{\mu(\text{it}-1)} = 1 \land x_{i,0}^{\mu(\text{it}-2)} = 1 \land x_{i,1}^\mu = 1 \land x_{i,1}^{\mu(\text{it}-1)} = 0 \land x_{i,1}^{\mu(\text{it}-2)} = 0$}
            \State $\Delta \mu_i \gets \Delta \mu_i/2$
        \ElsIf{$x_{i,0}^\mu = 1 \land x_{i,0}^{\mu(\text{it}-1)} = 0 \land x_{i,0}^{\mu(\text{it}-2)} = 0 \land x_{i,1}^\mu = 0 \land x_{i,1}^{\mu(\text{it}-1)} = 1 \land x_{i,1}^{\mu(\text{it}-2)} = 1$}
            \State $\Delta \mu_i \gets \Delta \mu_i/2$
        \ElsIf{$x_{i,0}^\mu = 0 \land x_{i,0}^{\mu(\text{it}-1)} = 1 \land x_{i,0}^{\mu(\text{it}-2)} = 0 \land x_{i,1}^\mu = 1 \land x_{i,1}^{\mu(\text{it}-1)} = 0 \land x_{i,1}^{\mu(\text{it}-2)} = 1$}
            \State $\Delta \mu_i \gets \Delta \mu_i/2$
        \ElsIf{$x_{i,0}^\mu = 1 \land x_{i,0}^{\mu(\text{it}-1)} = 0 \land x_{i,0}^{\mu(\text{it}-2)} = 1 \land x_{i,1}^\mu = 0 \land x_{i,1}^{\mu(\text{it}-1)} = 1 \land x_{i,1}^{\mu(\text{it}-2)} = 0$}
            \State $\Delta \mu_i \gets \Delta \mu_i/2$
        \EndIf
        \State Ensure $\underline{\Delta \mu} < \Delta \mu_i < \overline{\Delta \mu_i}$
    \EndFor
\EndFor
\end{algorithmic}
\end{algorithm}

\section{Results} \label{sec:result}
Four annealers are employed: QA, HA, DAv3, and QIIO. The number of readout samples per execution is fixed across all solvers, ranging from 2,000 to 100,000, and determined by the test case size. A 10-second time limit is assigned to HA. Minor embedding is employed for QA, with the chain strength selected to minimize chain breaks while preserving the relative importance of the original QUBO coefficients. The embedding target graph exhibits sparsity, mirroring the inherently low connectivity structure typical of power systems. For DAv3 and QIIO, the runtime configuration includes a 10-second time limit, 64-bit scaling precision, and a total optimization timeout of 3,600 seconds. \cref{tab:solvers} summarizes the capability of the solvers to converge to a solution within the residual tolerance of $1 \times 10^{-2}$ ${\text{MW}^2 + \text{MVAR}^2}\over 2$. For QA and HA, convergence indicates that an embedding has been found and that the solution is within the accepted residual tolerance. For DAv3 and QIIO, convergence implies that the solution is obtained within the same tolerance.

According to \cref{tab:solvers}, QIIO is capable of handling all test case sizes, including the 1354-bus test case. DAv3, on the other hand, is limited to test cases up to the 89-bus test case due to the constraints on the number of available binary variables. QA and HA are even more restricted and apply only to small test cases. The difference between DAv3 and QIIO lies in their hardware configurations: DAv3 provides a fully connected block of approximately 8,192 binary variables, while QIIO offers multiple fully connected blocks capable of managing up to 100,000 binary variables. This expanded capacity makes QIIO suitable for large-scale problems. Therefore, all further experiments in this paper are conducted using QIIO. Note that the experiments include test cases ranging from 4- to 1354-bus test cases from the Power System Test Cases library\footnote{\url{https://pandapower.readthedocs.io/en/v2.2.2}}. The results for the 4-, 9-, and 14-bus test cases are presented in \cite{kaseb2024power}, and the results for 5-, 6-, and 15-bus test cases are presented in \cite{kaseb2025combinatorial}.

\begin{table}[h] 
\caption{Performance comparison of different solvers under a user-defined residual of $1 \times 10^{-2}$ ${\text{MW}^2 + \text{MVAR}^2}\over 2$.} \label{tab:solvers}
\centering
\renewcommand{\arraystretch}{1.5}
    \begin{tabular}{lcccc}
        \toprule
        Test Case & QA & HA & DAv3 & QIIO \\ [1pt]
        \midrule
        4-bus    & \checkmark & \checkmark & \checkmark & \checkmark \\ 
        5-bus    & \checkmark & \checkmark & \checkmark & \checkmark \\
        6-bus    & \checkmark & \checkmark & \checkmark & \checkmark \\ 
        9-bus    & \checkmark & \checkmark & \checkmark & \checkmark \\ 
        14-bus   &            & \checkmark & \checkmark & \checkmark \\ 
        15-bus   &            &            & \checkmark & \checkmark \\ 
        30-bus   &            &            & \checkmark & \checkmark \\ 
        39-bus   &            &            & \checkmark & \checkmark \\ 
        57-bus   &            &            & \checkmark & \checkmark \\ 
        89-bus   &            &            & \checkmark & \checkmark \\ 
        118-bus  &            &            &            & \checkmark \\ 
        145-bus  &            &            &            & \checkmark$^*$ \\ 
        200-bus  &            &            &            & \checkmark$^*$ \\ 
        300-bus  &            &            &            & \checkmark$^*$ \\ 
        1354-bus &            &            &            & \checkmark$^*$ \\ 
        \bottomrule
        \multicolumn{5}{l}{\parbox{0.35\textwidth}{$^*$Experiment for the 145-, 200-, 300-, and 1354-bus test case is performed for a residual of $1 \times 10^{0}$ ${\text{MW}^2 + \text{MVAR}^2}\over 2$ due to the limited computational time.}}
    \end{tabular}
\end{table}

The performance of AQPF and AQOPF is compared with NR provided by \emph{pandapower} \cite{thurner2018pandapower} in \cref{subsec:partitioned}. These comparisons are performed based on the 118-bus test case\footnote{\url{https://icseg.iti.illinois.edu/ieee-118-bus-system/}}, which includes 118 buses, 173 branches, 13 transformers, 53 generators, and 54 loads. In \cref{subsec:ill-conditioned}, the performance of AQPF and AQOPF under ill-conditioned scenarios is evaluated for the 118-bus test case. In \cref{subsec:scalability}, a scalability analysis is presented for QIIO that demonstrates its ability to handle increasingly larger test cases. 

\begin{figure}[!htbp]
\centering 
\includegraphics[width=6.4in]{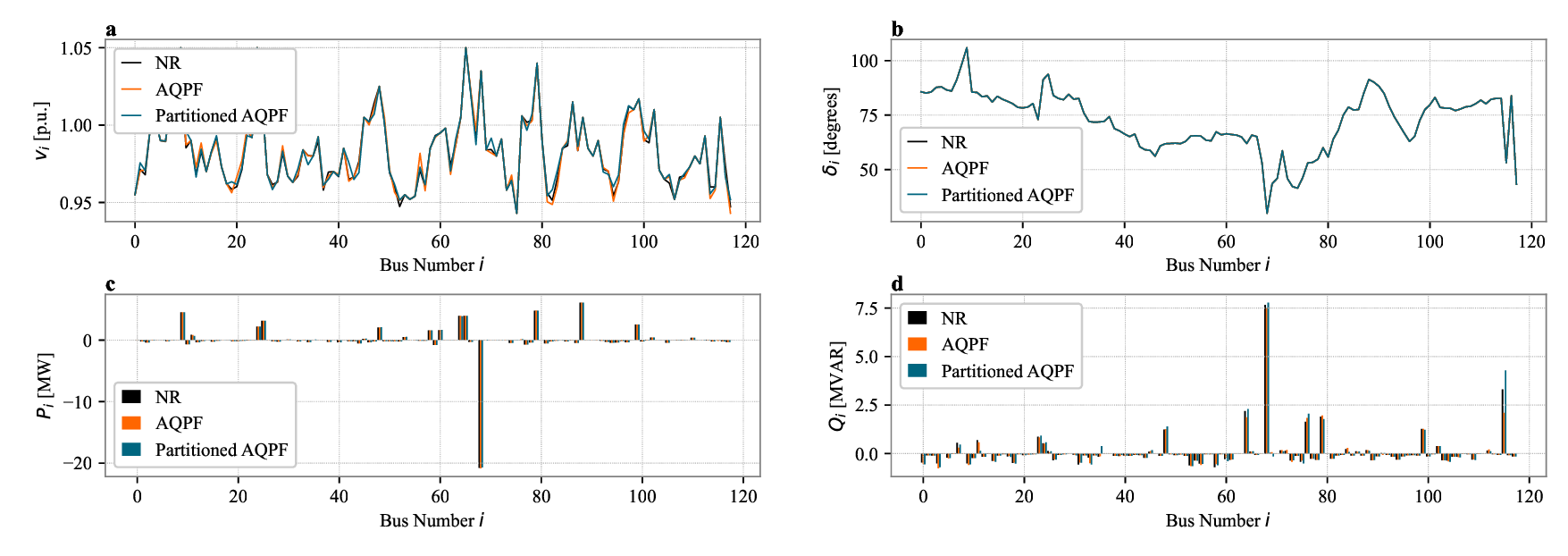}
\caption{Performance comparison of the NR method, AQPF, and partitioned AQPF for the 118-bus test case using QIIO. The results are shown for (a) voltage magnitude $V_i\;\text{(p.u.)}$, (b) voltage phase angle $\delta_i\;\text{(degrees)}$, (c) net active power $P_i\;\text{(MW)}$, and (d) net reactive power $Q_i\;\text{(MVAR)}$.} \label{fig:partitioned-aqpf}
\end{figure}

\begin{figure}[!htbp]
\centering 
\includegraphics[width=6.4in]{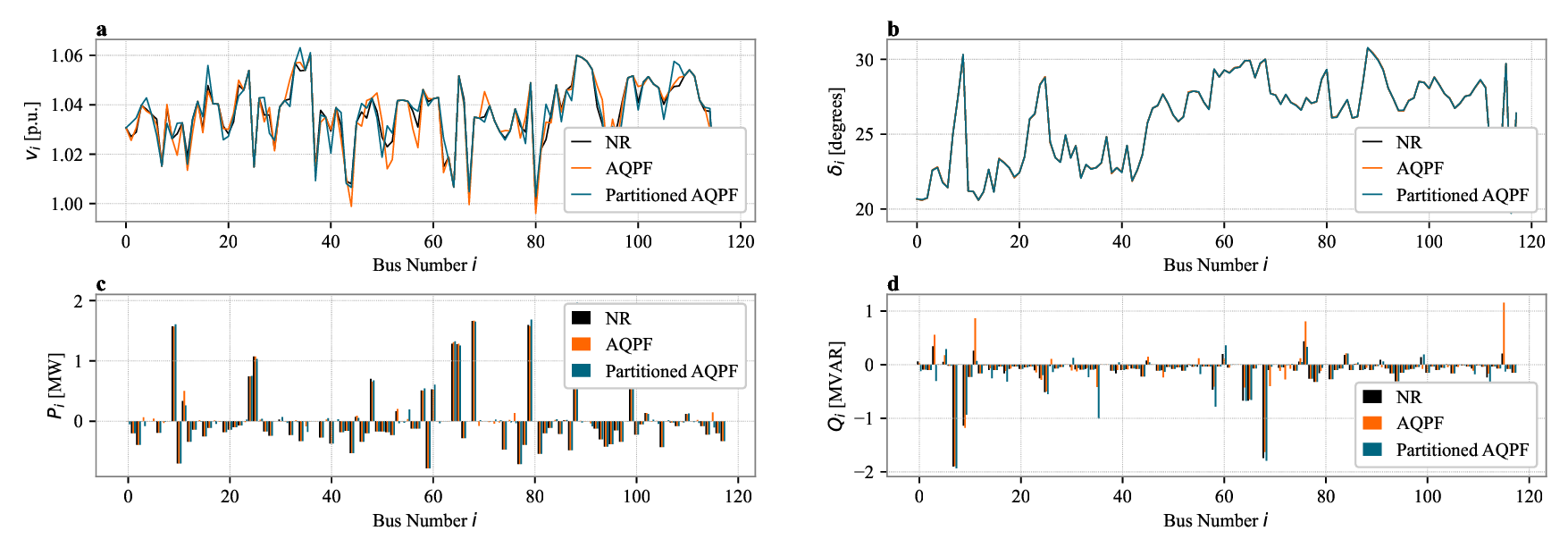}
\caption{Performance comparison of the NR method, AQOPF, and partitioned AQOPF for the 118-bus test case using QIIO. The results are shown for (a) voltage magnitude $V_i\;\text{(p.u.)}$, (b) voltage phase angle $\delta_i\;\text{(degrees)}$, (c) net active power $P_i\;\text{(MW)}$, and (d) net reactive power $Q_i\;\text{(MVAR)}$.} \label{fig:partitioned-aqopf}
\end{figure}

\subsection{Partitioned Formulation} \label{subsec:partitioned}
To assess the performance of the AQPF and AQOPF algorithms, their results, both in the full and partitioned formulations, are compared against NR for the 118-bus test case. The partitioned formulation is derived by excluding a subset of buses, denoted as $\mathbb{P}$, from the problem Hamiltonian in each iteration. Specifically, in each iteration, a set of 24 buses (approximately $20\%$ of the total buses) is randomly selected and removed from the computation. For PF analysis, this modification is applied in \cref{alg:aqpf} by excluding $\mathbb{P}$ from the summation term in \cref{eq:H_obj} at line 19, yielding:
\begin{align} \label{eq:H(118)}
    H_{\text{part}}(\vec{x}) = \sum_{i=1 \atop i \notin \mathbb{P} }^{N=118} (P_i - P_i^G + P_i^D)^2 + (Q_i - Q_i^G + Q_i^D)^2.
\end{align}
For OPF, the same subset $\mathbb{P}$ is excluded from the objective function in \cref{eq:final-hamiltonian}. This partitioned formulation reduces the problem size per iteration.

\subsubsection{Partitioned AQPF} \label{subsubsec:partitioned-aqpf} For PF analysis, the partitioned formulation in \cref{eq:H(118)} results in a total of $7105$ binary variables (including $472$ base variables and auxiliary variables, which are introduced to reduce higher-order terms), reducing the variable count by $1775$ compared to the full AQPF algorithm. With a residual threshold of $1 \times 10^{-2}$ ${\text{MW}^2 + \text{MVAR}^2}\over 2$, the mean square error (MSE) for the net active power $P_i$ between AQPF and NR is $4.28 \times 10^{-4}\;\text{MW}^2$, while for partitioned AQPF, it is $8.06 \times 10^{-4}\;\text{MW}^2$. 
Similarly, the MSE for the net reactive power $Q_i$ between AQPF and NR is $1.65 \times 10^{-2}\;\text{MVAR}^2$, while for partitioned AQPF, it is $1.8 \times 10^{-2}\;\text{MVAR}^2$. These results, illustrated in \cref{fig:partitioned-aqpf}, indicate that AQPF closely matches NR, and the partitioned formulation does not significantly degrade solution quality. Further improvements can be achieved by reducing the residual threshold. 

In terms of computational performance, AQPF requires $272.98\;\text{s}$ for compilation, compared to $208.34\;\text{s}$ for partitioned AQPF. The average time per iteration is $251.24\;\text{s}$ for AQPF and $217.61\;\text{s}$ for partitioned AQPF. Note that the total computational time includes communication overhead between the classical and quantum-inspired hardware. 

\subsubsection{Partitioned AQOPF} \label{subsubsec:partitioned-aqopf} For OPF, the partitioned formulation reduces the number of binary variables (including $472$ base variables, auxiliary variables, which are introduced to reduce higher-order terms, and slack variables, which are introduced to integrate the constraints) to $8459$, which represents a reduction of $2189$ binary variables compared to the full formulation in \cref{eq:final-hamiltonian}. With a residual of $1 \times 10^{-2}$ ${\text{MW}^2 + \text{MVAR}^2}\over 2$, the MSE for the net active power $P_i$ between AQOPF and NR is $8.89 \times 10^{-4}\;\text{MW}^2$, while for partitioned AQOPF, it is $1.57 \times 10^{-3}\;\text{MW}^2$. 
Similarly, the MSE for the net reactive power $Q_i$ between AQOPF and NR is $1.68 \times 10^{-2}\;\text{MVAR}^2$, while for partitioned AQOPF, it is $1.69 \times 10^{-2}\;\text{MVAR}^2$. 
The results are presented in \cref{fig:partitioned-aqopf}. 

The compilation times for AQOPF and partitioned AQOPF are $408.35\;\text{s}$ and $318.67\;\text{s}$, respectively. The time per iteration is $384.15\;\text{s}$ for AQOPF and $302.82\;\text{s}$ for partitioned AQOPF. These results indicate that the partitioned formulation can reduce computational time by up to $20\%$ while maintaining equivalent solution accuracy.

\subsection{Handling Ill-Conditioned Cases} \label{subsec:ill-conditioned}

\begin{table}[h] 
\caption{Comparison of AQPF and AQOPF under well-conditioned and ill-conditioned scenarios for the 118-bus test case.} \label{tab:ill-conditioned}
\centering
\renewcommand{\arraystretch}{1.5}
    \begin{tabular}{lcccccc}
        \toprule
         Scenario & \multicolumn{3}{c}{\textbf{AQPF}} & \multicolumn{3}{c}{\textbf{AQOPF}} \\ [1pt]
        \cline{2-7}
         & Residual & Mismatch & Mismatch & Residual & Mismatch & Mismatch \\ [0.2pt]
          & ${\text{MW}^2 + \text{MVAR}^2}\over 2$ & MW$^2$ & MVAR$^2$ & ${\text{MW}^2 + \text{MVAR}^2}\over 2$ & MW$^2$ & MVAR$^2$ \\ [1pt]
        \midrule
        Well-conditioned case    & $8.47 \times 10^{-3}$ & $4.28 \times 10^{-4}$ & $1.65 \times 10^{-2}$ & $8.85 \times 10^{-3}$ & $8.89 \times 10^{-4}$ & $1.68 \times 10^{-2}$ \\
        Ill-conditioned case 1   & $8.73 \times 10^{-3}$ & $5.75 \times 10^{-4}$ & $1.68 \times 10^{-2}$ & $8.59 \times 10^{-3}$ & $8.67 \times 10^{-4}$ & $1.63 \times 10^{-2}$ \\
        Ill-conditioned case 2   & $8.92 \times 10^{-3}$ & $5.24 \times 10^{-4}$ & $1.73 \times 10^{-2}$ & $8.87 \times 10^{-3}$ & $7.23 \times 10^{-4}$ & $1.71 \times 10^{-2}$ \\
        \bottomrule
    \end{tabular}
\end{table}

To simulate challenging operating conditions, two ill-conditioned cases are introduced for the 118-bus test case by modifying system parameters as follows:
\begin{enumerate}
    \item Load demands at a subset of \emph{PQ} buses are increased to create stress conditions that exceed the system's typical operating limits (Ill-conditioned case 1).
    \item The resistance ($R$) values of transmission lines connecting a subset of buses are increased, while the reactance ($X$) values remain unchanged. This results in significantly higher $R/X$ ratios and increasing network losses (Ill-conditioned case 2).
\end{enumerate}
Under these ill-conditioned scenarios, the NR solver in \emph{pandapower} diverges for PF. For OPF, numerical instability leads to infeasible solutions or failure to converge with optimization constraints remaining unmet. On the other hand, AQPF and AQOPF successfully solve both PF and OPF problems under the same ill-conditioned scenarios. The results are summarized in \cref{tab:ill-conditioned}. 

The capability of AQPF and AQOPF to handle ill-conditioned cases is attributed to their enhanced numerical stability; that is, their combinatorial formulation avoids reliance on traditional linearization techniques and hence mitigates issues caused by near-singular Jacobians. On the other hand, by iteratively refining increments/decrements, the algorithms ensure steady progress toward feasible solutions. The ability of Ising machines to handle a large number of binary decision variables also enables efficient searching of the solution space.

To further investigate the robustness of AQPF under ill-conditioned scenarios, the experiment is extended to include the partitioned AQPF formulation, as defined in \cref{eq:H(118)}, to solve PF analysis for the Ill-conditioned case 1. The results show that the partitioned AQPF formulation successfully solves the PF problem even under these challenging conditions. Specifically, the residual is $8.83 \times 10^{-3}$ ${\text{MW}^2 + \text{MVAR}^2}\over 2$, and the active and reactive power mismatches are $5.42 \times 10^{-4}\;\text{MW}^2$ and $1.49 \times 10^{-2}\;\text{MVAR}^2$, respectively. These results underscore the partitioned AQPF approach's ability to maintain computational efficiency without sacrificing solution accuracy in ill-conditioned scenarios.

\subsection{Scalability} \label{subsec:scalability}
The scalability of the AQPF and AQOPF algorithms is demonstrated by their ability to handle larger sets of binary variables using QIIO, as confirmed by experiments on test cases of varying sizes. The results are summarized in \cref{tab:aqpf-scalability} and \cref{tab:aqopf-scalability}, respectively, presenting the number of binary variables (-), the time needed to compile the QUBO formulation (s), the time elapsed per iteration (s), and the residual (${\text{MW}^2 + \text{MVAR}^2}\over 2$). Accordingly, QIIO can handle the number of binary variables required to solve PF up to the 1354-bus test case and OPF up to the 300-bus test case. However, due to the high computational complexity associated with the larger test cases, only a limited number of iterations are performed for them with a residual threshold of $1 \times 10^{0}$ ${\text{MW}^2 + \text{MVAR}^2}\over 2$, serving as a proof of concept. 

\begin{table}[h] 
\caption{Scalability of AQPF for different test cases.} \label{tab:aqpf-scalability}
\centering
\renewcommand{\arraystretch}{1.5}
    \begin{tabular}{lcccc}
        \toprule
        Test Case & No. of & Compile & Time per & Residual \\ [0.1pt]
        & Variables [-] & Time s & Iteration s & ${\text{MW}^2 + \text{MVAR}^2}\over 2$\\ [1pt]
        \midrule
        9-bus    & $249$    & $0.27$      & $0.93$      & $8.47 \times 10^{-3}$ \\ 
        14-bus   & $494$    & $39.62$     & $41.81$     & $8.11 \times 10^{-3}$ \\ 
        30-bus   & $1,183$  & $40.35$     & $40.91$     & $8.52 \times 10^{-3}$ \\ 
        57-bus   & $3,873$  & $90.05$     & $65.81$     & $8.57 \times 10^{-3}$ \\ 
        118-bus  & $8,881$  & $272.98$    & $251.24$    & $8.47 \times 10^{-3}$ \\ 
        200-bus  & $15,098$ & $238.22$    & $231.07$    & $9.38 \times 10^{-1}$ \\ 
        300-bus  & $23,577$ & $896.38$    & $926.64$    & $8.72 \times 10^{-1}$ \\ 
        1354-bus & $99,509$ & $43,151.54$ & $41,745.91$ & $9.67 \times 10^{-1}$ \\ 
        \bottomrule
    \end{tabular}
\end{table}

\begin{table}[h] 
\caption{Scalability of AQOPF for different test cases.} \label{tab:aqopf-scalability}
\centering
\renewcommand{\arraystretch}{1.5}
    \begin{tabular}{lcccc}
        \toprule
        Test Case & No. of & Compile & Time per & Residual \\ [0.1pt]
        & Variables [-] & Time s & Iteration s & ${\text{MW}^2 + \text{MVAR}^2}\over 2$\\ [1pt]
        \midrule
        9-bus    & $331$    & $0.42$      & $1.81$     & $8.59 \times 10^{-3}$ \\
        14-bus   & $692$    & $47.52$     & $48.51$    & $8.82 \times 10^{-3}$ \\
        30-bus   & $1.537$  & $52.11$     & $53.78$    & $8.73 \times 10^{-3}$ \\
        57-bus   & $6,021$  & $108.34$    & $96.29$    & $8.69 \times 10^{-3}$ \\ 
        118-bus  & $10,648$ & $408.35$    & $384.15$   & $8.85 \times 10^{-3}$ \\ 
        200-bus  & $18,571$ & $468.08$    & $406.65$   & $9.84 \times 10^{-1}$ \\ 
        300-bus  & $29,498$ & $1,659.12$  & $1,395.87$ & $9.79 \times 10^{-1}$ \\ 
        \bottomrule
    \end{tabular}
\end{table}
 
\section{Discussion} \label{sec:discussion}
The results presented in this study should be interpreted within the broader context of algorithm development for emerging quantum computing paradigms. The intention of the proposed AQPF and AQOPF algorithms is not to compete with or replace established classical PF and OPF solvers, which have benefited from decades of research and remain highly effective for most practical applications. Instead, this study focuses on developing algorithms that could be useful as quantum hardware continues to mature. The proposed discrete combinatorial optimization approach provides an alternative mathematical representation of PF and OPF equations, enabling their implementation on Ising machines. At the same time, this representation leads to NP-hard optimization problems, and this complexity aligns with the capabilities targeted by quantum and digital annealers. Although current devices impose limitations on solution precision and problem scale, the experiments demonstrate that the methodology is computationally scalable and compatible with multiple annealing platforms. The following observations can be made:
\begin{itemize}
    \item Due to the high computational cost, the experiments are conducted with a threshold of $1 \times 10^{-2}$ ${\text{MW}^2 + \text{MVAR}^2}\over 2$. Given additional computational time, the results could be further refined to obtain higher accuracy.
    \item QA and HA are constrained by embedding challenges and available qubit count, making them unsuitable for large test cases. While they successfully solve small-scale problems, their applicability to real-world power systems remains limited without improvements in embedding techniques, hardware connectivity, and qubit counts.
    \item QIIO offers a remarkable capability to handle up to 100,000 binary variables, making it an excellent platform for solving large-scale power system problems.
    \item A more efficient update scheme can significantly enhance the performance of the proposed algorithms by minimizing computational overhead while maintaining accuracy. Future work can therefore improve the update scheme to reduce the number of iterations required for convergence.
    \item The AQPF algorithm can be generalized as a “model of models” approach that is executable on Ising machines. This capability opens avenues for tackling a wide range of problems beyond power systems, particularly those that are computationally challenging for classical solvers.
\end{itemize}

\section{Conclusion} \label{sec:conclusion}
This study further investigates the PF and OPF problem reformulation as discrete combinatorial optimization problems that can be executed on Ising machines, including quantum and quantum-inspired hardware. Building upon the previously proposed Adiabatic Quantum Power Flow (AQPF) algorithm, this study introduces the Adiabatic Quantum Optimal Power Flow (AQOPF) algorithm. It evaluates the feasibility of both algorithms across a wide range of test cases. In addition, the computational efficiency and scalability of the proposed algorithms are highlighted. The findings confirm that the AQPF and AQOPF algorithms consistently converge across various test cases, including complex and ill-conditioned cases where conventional solvers often struggle. The results also suggest that QIIO offers a compelling alternative to classical solvers, particularly for problems with high-dimensional, combinatorial structures. This study, therefore, marks a significant step toward integrating Ising machines into power system applications and paves the way for enhanced efficiency, scalability, and robustness in solving large-scale optimization problems. 

\section*{Acknowledgment}
This work was carried out within the DATALESs project (project no. 482.20.602), financially supported by the Netherlands Organization for Scientific Research (NWO) and the National Natural Science Foundation of China (NSFC).
The authors gratefully acknowledge Heinz Wilkening and the service contract "Quantum Computing for Load Flow" (contract no. 690523) with the European Commission Directorate-General Joint Research Centre (EC DG JRC). Furthermore, the authors thank the J\"ulich Supercomputing Centre for allocating computing resources on the D-Wave Advantage\texttrademark\, System JUPSI via the J\"ulich UNified Infrastructure for Quantum computing (JUNIQ). This research was supported by the Center of Excellence RAISE, funded under the European Union’s Horizon 2020 Research and Innovation Framework Programme H2020-INFRAEDI-2019-1 (agreement no. 951733). 
The authors also thank Fujitsu Technology Solutions for granting access to the QIIO software\footnote{\url{https://en-portal.research.global.fujitsu.com/kozuchi}}, with special appreciation to Markus Kirsch and Matthieu Parizy for their assistance and for providing customized extensions to the DADK Python package.

\printcredits

\bibliographystyle{unsrt}
\bibliography{cas-refs}

\end{document}